\documentclass[reprint,showpacs,preprintnumbers,amsmath,amssymb,aip]{revtex4-1}
\usepackage{graphicx,wasysym}
\usepackage{dcolumn}
\usepackage{bm}
\usepackage{bbm}
\usepackage{natbib}
\usepackage{tabularx}
\usepackage{color}
\usepackage{hyperref}

\newcommand{\kB}{k_\mathrm{B}}

\begin{document}
\title{Learning intermolecular forces at liquid-vapor interfaces}
\author{Samuel P. Niblett} 
\affiliation{Department of Chemistry, University of California, Berkeley CA 94609 \looseness=-1}
\affiliation{Materials Science Division, Lawrence Berkeley National Laboratory, Berkeley, CA 94609 \looseness=-1}

\author{Mirza Galib} 
\affiliation{Department of Chemistry, University of California, Berkeley CA 94609 \looseness=-1}

\author{David T. Limmer} 
\email{dlimmer@berkeley.edu}
\affiliation{Department of Chemistry, University of California, Berkeley CA 94609 \looseness=-1}
\affiliation{Materials Science Division, Lawrence Berkeley National Laboratory, Berkeley, CA 94609 \looseness=-1}
\affiliation{Chemical Science Division, Lawrence Berkeley National Laboratory, Berkeley, CA 94609\looseness=-1}
\affiliation{Kavli Energy NanoScience Institute, Berkeley, CA 94609 \looseness=-1}

\date{\today}
\begin{abstract}
By adopting a perspective informed by contemporary liquid state theory, we consider how to train an artificial neural network potential to describe inhomogeneous, disordered systems. We find that neural network potentials based on local representations of atomic environments are capable of describing some properties of liquid-vapor interfaces, but typically fail for properties that depend on unbalanced long-ranged interactions which build up in the presence of broken translation symmetry. These same interactions cancel in the translationally invariant bulk, allowing local neural network potentials to describe bulk properties correctly. By incorporating explicit models of the slowly-varying long-ranged interactions and training neural networks only on the short ranged components, we can arrive at potentials that robustly recover interfacial properties. We find that local neural network models can sometimes approximate a local molecular field potential to correct for the truncated interactions, but this behavior is variable and hard to learn. Generally, we find that models with explicit electrostatics are easier to train and have higher accuracy. We demonstrate this perspective in a simple model of an asymmetric dipolar fluid where the exact long-ranged interaction is known, and in an \emph{ab initio} water model where it is approximated. 
\end{abstract}
\maketitle

\section{Introduction}
Machine learning based forcefields have significantly broadened the scope of molecular simulations.\cite{behler2016perspective,noe2020machine,bartok2017machine} Supervised learning techniques have been used to refine the parameterization of traditional physically motivated potentials,\cite{cisneros2016modeling,li2017machine,unke2019physnet,cubuk2017representations} and enabled those employing artificial neural network (ANN) representations.\cite{Zhang2018NeurIPS,behler2007generalized,behler2011atom,Wang2018}  ANN potentials in particular have demonstrated their utility in the prediction of material properties,\cite{Csanyi2017,marchand2020machine,bischak2020liquid} and in the accurate calculation of thermodynamic\cite{behler2008pressure,Cheng2018,morawietz2016van,bonati2018silicon} and kinetic properties.\cite{yang2021using,Galib21,kaser2020isomerization,liu2018constructing} While there have been significant advances in the chemical complexity of systems studied with ANN potentials, as they are flexible enough describe reactivity,\cite{behler2017first} the overwhelming majority of studies have employed relatively simple bulk environments, with a few notable exceptions.\cite{natarajan2016neural,andrade2020free,Wohlfahrt2020,Galib21,Yue2020} A barrier preventing the extension of ANN potentials into more complex environments is that they are truncated at a finite interaction distance, which foundational work in liquid state theory suggests can lead to inaccuracies in systems lacking translational invariance.\cite{weeks1971role,lum1999hydrophobicity,weeks1995self} Here we employ a perspective based on the local molecular field theory to rationalize the failings of truncated ANN potentials in nonuniform systems and to develop a representation and training procedure that admits their extension to liquid-vapor interfaces. 

Local molecular field (LMF) theory was introduced by Weeks and collaborators to understand the structure and thermodynamic properties of nonuniform, disordered systems.\cite{weeks1995self,weeks2002connecting} The LMF theory uses the Yvon-Born Green hierarchy\cite{hansen1990theory} to demonstrate that a system with slowly varying long-ranged interactions, like those associated with electrostatic or dispersion forces, can be mapped to a system with short-ranged intermolecular interactions in an effective external field. In a bulk environment, translational invariance dictates that the effective external field is a constant and therefore inconsequential. At an interface, this external field corrects for the unbalanced interactions in the truncated effective model. This perspective has been used analytically in theories of solvation, and computationally as a means to model electrostatic interactions without traditional Ewald sums.\cite{rodgers2008local,remsing2011deconstructing,rodgers2006attraction,rodgers2008local,remsing2014role,remsing2013dissecting,remsing2016role,cox2020dielectric}

ANN potentials are typically formulated with mathematical representations of a system, so-called symmetry functions, that are short-ranged,\cite{behler2007generalized} yet aim to approximate intermolecular potentials that are often long-ranged.  As such, it is natural to use LMF theory as a lens to understand the limitations and successes of truncated ANN potentials. This is the perspective we adopt here. Consistent with LMF theory, we find that bulk properties are insensitive to the description of long-ranged interactions, while some interfacial properties, like the interfacial polarization, are particularly sensitive. LMF theory also suggests a means of ameliorating these mixed successes by incorporating the slowly-varying part of the electrostatic interactions in extended ANN potentials.

The limitations of ANN potentials employing only short-ranged interactions have been discussed extensively, and a number of procedures have been proposed to circumvent them.\cite{Grisafi2019,Deng2019,Yue2020,bartok2010gaussian,artrith2011high,yao2018tensormol} The short-ranged character of ANN potentials results from the symmetry functions using a cutoff distance, so the most direct route to incorporate long-ranged interactions is to include non-local features in those molecular representations.\cite{Grisafi2019} However, a more common approach involves explicitly evaluating additional long-ranged forces.\cite{Deng2019,Yue2020,bartok2010gaussian,artrith2011high,yao2018tensormol} In this framework, the neural network is trained to reproduce a truncated version of the target potential, for example in a classical model the ANN might be trained only on the van der Waals component of the energy. During simulation, the interactions predicted by the ANN are combined with an explicit calculation of the remaining long-ranged components. The advantage of this method is that long-ranged interactions are included directly, while the disadvantages are the cost and the ambiguity in selecting the models used to evaluate the long-ranged forces. We adopt this latter approach, employing an explicit description of the electrostatic interactions both in systems where the true description is known and in cases where it is not and we must approximate it.  The addition of explicit long-ranged forces into the ANN models increases the overall accuracy of the models, decreases the difficulty training them, and consistently recovers the interfacial properties of the reference model. Further, we find that the ANNs are not very sensitive to the specific auxiliary long-ranged force model employed. In the following these findings are explored within the context of a simple model of an asymmetric dipolar fluid, and in an \emph{ab initio} model of liquid water.

\section{Reference Models}
\begin{figure}[t]
    \includegraphics[width=8.5cm]{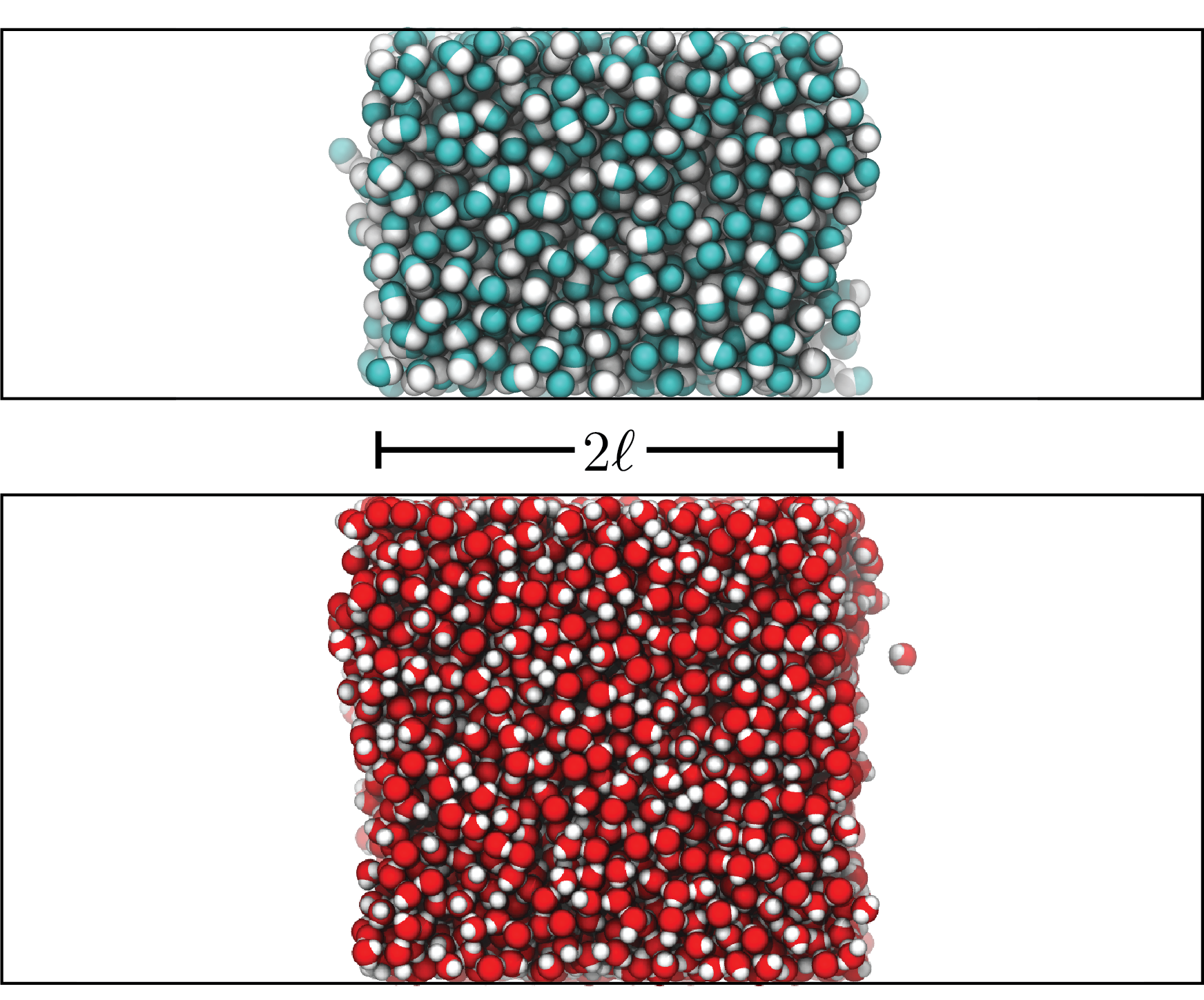}
  \caption{Characteristic snapshot of the slab geometries of the dipole fluid (top) and water (bottom), showing also the lengthscale $\ell$ which indicates the half width of the liquid slab.}
  \label{fig:Model_geom}
\end{figure}

In order to understand the fundamental limitations of truncated ANN potentials and how to move beyond them, we study ANN potentials for two systems: a fixed charge forcefield model and an \emph{ab initio} model. In all subsequent discussion, we will use ``reference model'' to describe the target potential that an ANN is trained to reproduce. For the forcefield, we can compute the reference values of ensemble-averaged quantities precisely, in order to validate collective properties of the ANN system. Further, in this reference model the form of the long-ranged force is known by construction so that we can assess its impact in a controlled manner. We also study an \emph{ab initio} model based on density functional theory, analogous to most of the current efforts at parameterizing ANN potentials. In such a setting, the exact form of the long-ranged interactions is not known, allowing us to test the sensitivity to our choice of interaction model. We consider a relatively simple heterogeneous environment, a liquid in contact with its vapor using a standard slab geometry, shown in Fig.~\ref{fig:Model_geom}. The width of the slab, equal to the length between the two Gibbs dividing surfaces, is defined as $2\ell$ and is used throughout to scale the $z$ axis perpendicular to the interface. This scaling allows us to compare systems with slightly different densities. All ANN calculations are done in LAMMPS\cite{PLIMPTON19951} using the DeePMD extension.\cite{Wang2018}

\subsection{Asymmetric dipolar fluid} 
The forcefield to which we fitted an ANN is a single-component dipolar fluid constructed from a flexibly bonded dimer. This forcefield was adapted from a model of CO,\cite{Daub2014} made flexible to avoid difficulties in learning the rigid body constraints. The two sites of each dimer interact only through a harmonic bonding potential, with the form 
\begin{equation}
\label{Eq:bond}
U_\mathrm{B}(r_{AB}) = \frac{1}{2} k_\mathrm{bond} \left (r_{AB}-r_\mathrm{bond} \right )^2 
\end{equation}
where $r_{AB}$ is the displacement between the $A$ and $B$ sites of the dimer, held at an
equilibrium displacement distance $r_\mathrm{bond}=1.2782$\AA \, with force constant $k_\mathrm{bond}=100 \mathrm{kcal}/\mathrm{mol}$\,\AA$^2$. By virtue of the empirical forcefield, the short- and long-ranged interactions are known explicitly. These interactions are comprised of a Lennard-Jones potential on each site and compensating point charges. The Lennard-Jones potential for pairs $i-j$ takes the form
\begin{equation}
U_\mathrm{LJ}(r_{ij}) = 4 \epsilon \left [ \left ( \frac{\sigma_{ij}}{r_{ij}}\right )^{12} -\left ( \frac{\sigma_{ij}}{r_{ij}}\right )^{6} \right ]
\end{equation}
where $r_{ij}$ is the displacement between sites $i$ and $j$, and $\epsilon=0.08413 \,\mathrm{kcal/mol}$ sets the energy scale for the potential. The dimer is made asymmetric to break charge inversion symmetry so that the extended interface can support a net polarization. This is done by choosing the diameters of the $A$ and $B$ sites to be different, with $\sigma_{AA} = 2 \sigma_{BB}=2.18113$\AA, and $\sigma_{AB}=3\sigma_{BB}/2$.  The Lennard-Jones potential is truncated and shifted at $r_{\rm cut}=11.45$\AA \, and thus is considered a short ranged interaction. Finally, each site is charged and thus interacts electrostatically according to the pair potential
\begin{equation}
\label{Eq:C}
U_\mathrm{C}(r_{ij}) = \frac{q_i q_j}{4 \pi \varepsilon_0 r_{ij}},
\end{equation}
with site charges $q_A=-q_B= 0.21 q_e$ where $q_e$ is the charge of the electron and $\varepsilon_0$ is the permittivity of free space. The Coulomb potential is not truncated and is evaluated using a particle-particle mesh Ewald summation. The net neutrality and asymmetry of the dimer results in asymptotic dipole-dipole interactions at long distances, and the flexible bond endows it with a low frequency polarizability. The two sites are also treated as point masses with mass 28 amu. 

The liquid-vapor interface is studied at 60 K or a reduced temperature of $T^*=\kB T/\epsilon = 1.4$, which is below the critical point for the rigid model, reported to be $T_c^*=2.0$.  The temperature was controlled using a Langevin thermostat with time constant of 10 ps.  Simulations were initialized from a regular lattice with 1200 molecules or $N=2400$ atoms and equilibrated for 10ns with a constant pressure simulation at 1 atm using the Nose-Hoover algorithm.  The simulation box was then extended by 5 nm in the $+z$ and $-z$ directions, creating two liquid-vapor interfaces perpendicular to the $z$ axis and a 10 nm vacuum gap. The final simulation box dimensions were $L_x=L_y=2.157$nm, $L_z=15.808$nm. Training data for fitting the ANN were selected at 50ps intervals from a 100ns simulation at constant temperature and volume, allowing 1 ns for equilibration.

\subsection{revPBE-D3 water}
%
We have also trained a set of ANN models to reproduce a many body interatomic potential energy surface for water, computed by density functional theory. The primary data sets for the training were generated from \emph{ab initio} molecular dynamics simulations using the Gaussian Plane Wave implementation in CP2K.\cite{VandeVondele2005} All \emph{ab initio} molecular dynamics simulations were carried out using the revised version of the PBE functional\cite{zhang1998comment} along with empirical dispersion correction (Grimme D3).\cite{grimme2010consistent} We used a molopt-DZVP basis set and a plane wave cut-off of 1200\,Ry. This large cutoff was needed to converge the virial component of the pressure. The core electrons were described with the GTH pseudopotential.\cite{goedecker1996separable} 

The liquid-vapor interface is studied at 300\,K.  The temperature was controlled using a Langevin thermostat with time constant of 1\,ps.  Simulations were initialized from a regular lattice with 340 molecules or $N=1020$ atoms and equilibrated for 20 ps with a constant pressure simulation at 1 atm using the Nose-Hoover algorithm.  The simulation box was then extended by 1.5\,nm in the $+z$ and $-z$ directions, creating two liquid-vapor interfaces perpendicular to the $z$ axis and a 3nm vacuum gap. The final simulation box dimensions were $L_x=L_y=2.0$\,nm, $L_z=5.0$\,nm. Training data for fitting the ANN were selected at 0.005ps intervals from a 25\,ps simulation at constant temperature and volume, allowing 20\,ps for equilibration. A total of 5000 configurations were used in the training. 

\section{ANN training methods}

We employ the DeePMD scheme to evaluate and train our ANN potentials.\cite{Wang2018,zhang2018deep} As with almost all neural network potentials, DeePMD assigns atomic energies and forces using a local mathematical description of the environment.\cite{Zhang2018NeurIPS} The initial representation of atom $i$'s environment is given by a matrix $\boldsymbol{\mathcal{R}}^i \in \mathbbm{R}^{N_i\times 4}$ where $N_i$ is the number of atoms within $r_c=8$\AA \,of atom $i$. Each row of $\boldsymbol{\mathcal{R}}^i$ represents one neighbor and is given by $\mathcal{R}^{i}=\{s(r_{ji}),\hat{x}_{ji},\hat{y}_{ji},\hat{z}_{ji}\}$ where $\hat{x}_{ji}=s(r_{ji})x_{ji}/r_{ji}$ is a scaled unit vector pointing from atom $i$ to neighbor $j$. Analogous definitions hold for  $\hat{y}_{ji}$ and  $\hat{z}_{ji}$.  The scaling function $s(r_{ji})$ assigns greater weight to the nearest neighbors
\begin{equation}
  s(r_{ji}) = \begin{cases}
    \frac{1}{r_{ji}}, & r_{ji} < r_{cs} \\
    \frac{1}{r_{ji}}\{\frac{1}{2}\cos[\pi\frac{r_{ji}-r_{cs}}{r_c-r_{cs}}]+\frac{1}{2}\}, & r_{cs} < r_{ji} < r_{c} \\
    0, & r_{ji} > r_c.
  \end{cases}
  \label{eq:L2}
\end{equation}
The parameter $r_{cs}=7.9$\AA~ defines a switching radius beyond which a cosine function is used to take the scaling factor smoothly to zero at $r_c$. This smoothness guarantees differentiability of the scaling function, which is required for well-conditioned minimization and energy conserving dynamics.

Another requirement for reliable ANN potentials is that environment representations should be invariant to global symmetries of the system including translation, rotation and atomic permutation.\cite{behler2007generalized,glielmo2017accurate,grisafi2018symmetry} DeePMD achieves this invariance by converting $\boldsymbol{\mathcal{R}}^i$ to an invariant feature matrix $\mathcal{D}^i\in\mathbbm{R}^{M_1\times M_2}$ using an encoding neural network whose parameters are optimized simultaneously with the network representing the potential energy.\cite{Zhang2018NeurIPS}  We employed a 3-layer encoding network containing 25, 50 and 100 neurons in each layer, with a fourth axis layer containing 16 neurons. For the dipolar model, the main fitting network is also a three layer neural network containing 300 neurons per layer. For the \emph{ab initio} water model, the network is a three layer neural network containing 600 neurons per layer.

\subsection{Loss function for optimization}
During training, a loss function is minimized with respect to all neural network parameters using the Adam stochastic gradient descent algorithm.\cite{Adam} For the $n$th optimization step, the loss function is given by an average over a set of $N_{\rm batch}$ configurations, denoted $\mathbf{R}$,
\begin{equation}
 \mathcal{L}(n) = \frac{1}{N_{\rm batch}}\sum_\mathbf{R}^{N_{\rm batch}} L(\mathbf{R},n)
\end{equation}
with a step-dependent function
\begin{equation}
L(\mathbf{R},n)= p_e(n)\Delta U^2 + \frac{p_f(n)}{3N}\sum_i^{N} |\Delta\mathbf{F}_i|^2 + \frac{p_\xi(n)}{9N}||\Delta \Xi||^2 \, .
\end{equation}
Here $\Delta U(\mathbf{R}) = U_{\rm ANN}(\mathbf{R}) - U_{\rm ref}(\mathbf{R})$ is the prediction error of the ANN for the total configuration potential energy, $\Delta \mathbf{F}_i = \mathbf{F}_{i,{\rm ANN}}(\mathbf{R}) - \mathbf{F}_{i,{\rm ref}}(\mathbf{R})$ is the error for the force on atom $i$, and $\Delta\Xi = \Xi_{{\rm ANN}}(\mathbf{R}) - \Xi_{{\rm ref}}(\mathbf{R})$ is the elementwise error in the virial matrix, $\mathbf{\Xi}$=$-\sum_i\mathbf{R}_i\otimes\mathbf{F}_i/2$. The three prefactors $p_e, p_f$ and $p_\xi$ are weights that vary with optimization step number $n$ according to $p(n) = p(n_f)[1-r_l(n)/r_l(0)]+p(0)[r_l(n)/r_l(0)]$, where $r_l(n)$ is the learning rate and $p(n_f)$ is a specified final value. The learning rate (i.e.~the Adam optimizer step size) decreases during optimization following $r_l(n) = r_l(0)d_r^{(n/d_s)}$, where $d_r$ and $d_s$ are termed the decay rate and decay steps, respectively. In the limit $N_{\rm batch}$ equals the whole training set, $\mathcal{L}(n)$ becomes deterministic, but in general a subset of the total configurations are sampled randomly.

In addition to the expected representability problems with employing ANNs with truncated interactions to model interfaces, there is a practical training problem. The proportion of atoms representing interfacial environments is small, which can lead to under-learning of forces and higher prediction errors for interfacial atoms relative to those in bulk-like environments. We found that the stability of molecular dynamics and the accuracy of interfacial properties are particularly sensitive to the maximum error of the neural network, as well as to the absolute average error.
To mitigate this environment bias we have found it useful to introduce a weighting function to the loss function to homogenize high-error atomic environments. Specifically, we dress the squared force error
\begin{equation}
  \label{eq:error_switch}
 |\Delta \mathbf{F}_i|^2 \rightarrow |\Delta \mathbf{F}_i|^2\left[ w - \delta w \, {\rm tanh}\left(g -|\Delta\mathbf{F}_{i}|^2 /\bar{f}^2\right) \right ]
\end{equation}
that saturates at large forces. This weighting function is not appropriate for an initial training job, where most configurations have large force errors, but can be beneficial during active learning steps for model refinement (described in a subsequent section). We employed the weighting function when retraining ANN representations of the DFT water model. We found empirically that $w=502,~\delta w=500, \bar{f}=0.022 e\mathrm{V/}$\AA \,and $g=14$ works well in training these systems.

There are many tunable hyperparameters associated with the training algorithm. We have found that parameter choice rarely makes a large difference to the minimal loss function obtained during a training run, but it often does affect the number of steps required to reach this value. Moreover, the variance in training quality and model properties between ANNs trained using different hyperparameters is no greater than the variance between training processes with the same hyperparameters but different random seeds. Example hyperparameters for training the dipole forcefield model are shown in Table~\ref{tab:training_params}. 

\begin{table}[b]
  \begin{tabularx}{.45\textwidth}{c|c|c|c}
    Hyperparameter \quad & \quad Training\quad \quad& \quad Training \quad\quad& \quad Training\quad \quad \\
      & \quad step 1 \quad& \quad step 2 \quad& \quad step 3 \quad \quad \\
  \hline
  $N_{\rm batch}$ & 30 & 1 & 1  \\ 
  $l_r(0)$ & $10^{-3}$ & $10^{-3}$ & $10^{-4}$ \\
  $d_r$ & 0.999 &0.999 & 0.999 \\
  $d_s$ & 10 & 10 &1000\\
  $p_e(0)$ & 0.01 & 0.01 & 0.01\\
  $p_e(n_f)$ & 1.0 & 0.01& 0.01\\
  $p_f(0)$ & 100.0 &1.0 & 1.0\\
  $p_f(n_f)$ & 1.0 &1.0 & 1.0\\
  $p_\xi(0)$ & 1.0 & 0.0 & 0.0 \\
  $p_\xi (n_f)$ & 1.0& 0.0 & 0.0\\
  \hline
\end{tabularx}
\caption{\label{tab:training_params}One set of example hyperparameters for training the dipole model. Three training steps were performed, the first generates an initial model and the others refine it using additional data obtained through active learning protocols.}
\end{table}


\subsection{\label{sec:LRModels}Explicit Long-ranged ANN models}
In order to understand the impact of long-ranged forces on interfacial structure and thermodynamics, for each reference system we train one model using standard short-ranged ANNs and another supplemented with an explicit model of the long-ranged interaction. The former, which we refer to as ANN-SR for short-ranged interactions, is trained on the full potential energy function $U(\mathbf{R})$ of the reference model and its derivatives which provide information on the forces and virial. For the dipolar fluid model, the total potential consists of a sum over the explicit pairwise bonding, Lennard-Jones and Coulomb potentials. 
For the revPBE-D3 water model, the total potential energy derives from an expectation over the minimized electron density, and the forces are derived from the Hellman-Feynman theorem.\cite{VandeVondele2005} 

The model with explicit long-ranged interactions, which we refer to as ANN-LR for long-ranged interactions, is trained on only a piece of the reference potential energy and force, while an additional piece is treated with an empirical model. In contrast to previous work,\cite{Yue2020} we do not train the ANN-LR models on the difference between the total potential energy and the full electrostatic potential. Rather we subtract only the long-ranged, slowly varying interactions. This splitting is accomplished using a construction supplied by LMF theory, which defines the short ranged electrostatic potential as\cite{rodgers2006attraction}
\begin{equation}
\label{Eq:LMD}
U_{\rm SR}(\mathbf{R}) = U(\mathbf{R}) - \sum_{i<j} \frac{q_i q_j}{4 \pi \varepsilon_0 r_{ij}} {\rm erf}(\kappa r_{ij} ).
\end{equation}
Here the error function, ${\rm erf}(\kappa r_{ij})$, has a natural lengthscale $\kappa^{-1}$ and multiplies a pairwise Coulomb model to extract the long-ranged interactions. The ANN-LR models are trained to reproduce only $U_{\rm SR}(\mathbf{R})$ and its derivatives, while simulations are run using that forcefield combined with an exact treatment of the remainder. Several efficient methods exist for evaluating the remainder, and we employ the linear scaling particle particle mesh Ewald sum.\cite{hockney2021computer} For the dipolar fluid model, this procedure is equivalent to training an ANN on that part of the potential energy typically evaluated in real-space, including the bonding, Lennard-Jones, and part of the Coulomb potential, while still explicitly evaluating the reciprocal space part of the Coulomb potential. Implementation details of this procedure are briefly discussed in Appendix A.

While $U_\mathrm{C}(r_{ij})$ is known explicitly in the dipolar fluid model, for the revPBE-D3 water model, the long-ranged part of the interactions is not known. We model it as the long-ranged electrostatic force arising from a fixed Gaussian charge density centered on each atom. Below, we explore the sensitivity of this approach to the choice of Gaussian amplitude, but most of our results employ charges taken from the empirical SPC/E forcefield with hydrogen charge $q_H=0.4238q_e$.\cite{berendsen1987missing} 

In principle, the charges and $\kappa$ could be optimized alongside the ANN, but for simplicity we do not consider doing so here and instead employ fixed values. 
Within LMF theory, it is known that $\kappa$ must be chosen small enough that pair correlations are still well-described in the absence of the long-ranged potential.\cite{rodgers2008local}
Neural network potentials are capable of capturing these correlations on length scales smaller than the symmetry function cutoff $r_c$, so we expect that $\kappa^{-1}\leq r_c$ will be a suitable choice. In cases where the true charge distribution is not known, it will be desirable to minimise the role of the explicit charge model by choosing $\kappa$ as large as possible. We therefore consider the case $\kappa^{-1}=r_c=8.0$\AA.
We find that this choice gives good training results for both the dipole and \emph{ab initio} water models.

 \begin{figure}[t]
    \includegraphics[width=8.cm]{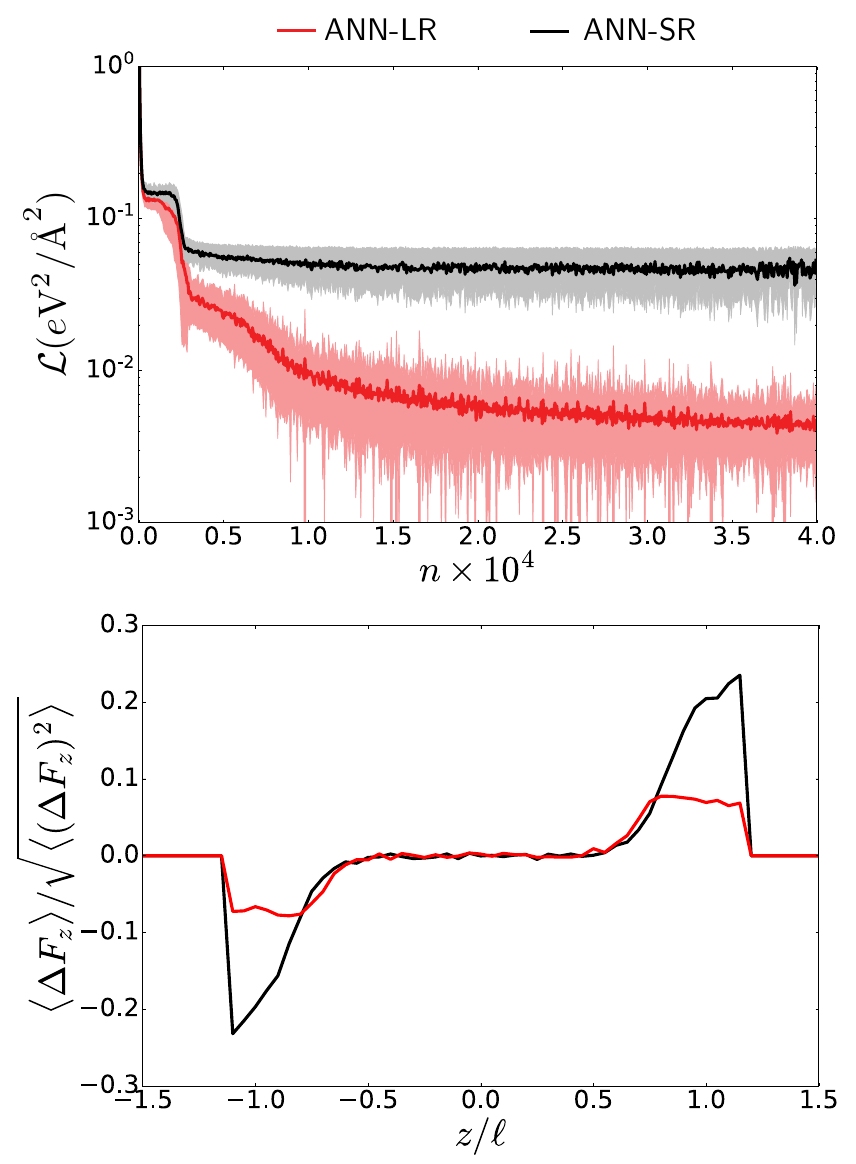}
  \caption{A) Ensemble-averaged learning curves for ANN models of the dipole fluid model. Dark lines correspond to the loss function including only the contribution from the force error. Shaded areas indicate the standard deviation of the set of models.  B) Average errors in $z$-force prediction as a function of atomic $z$ coordinate for ANN-SR and ANN-LR models, normalized by the root-mean squared error for each model. Configurations used in the ensemble average were drawn from the ANN training set used for the models. Both panels A and B are averaged over 20 training runs initialized with different random seeds, or their resultant models. The $z$ coordinates were normalised to the mean interface positions $\pm\ell$ before averaging over the models, and the quantity $\langle \Delta F_z \rangle$ is set to 0 for $z$ values where the A-type mass density is less than 2\% of the bulk liquid value.}
  \label{fig:ave_lcurve}
\end{figure}

 Our approach provides a clear separation between the short-ranged component of the energy that is representable by a ANN and the long-ranged component that is not. We find that this separation provides an accurate and efficient means of modeling extended interfaces with ANNs. Figure~\ref{fig:ave_lcurve}A) shows an ensemble of initial learning curves for ANN-SR and ANN-LR models of the dipole fluid, demonstrating that the latter attains a significantly lower loss function, leading to more accurate predictions of properties and greater consistency between models. The converged loss function of the ANN-LR model is approximately an order of magnitude smaller than that of the ANN-SR model. This difference is expected, since the ANN-SR attempts to fit correlations between short-ranged structure and long-ranged energies that are not easily represented by the truncated symmetry functions. 
 
 Figure~\ref{fig:ave_lcurve}B) plots the force error averaged over an equilibrium trajectory conditioned by the $z$ value of the atom and normalized by the mean squared error of each model. This figure illustrates that the residual error in the ANN-SR models is mostly due to the presence of the interface.  The errors in the bulk region of the slab are essentially the same in both models and very small, which indicates that  both models are able to learn the short-ranged part of the potential well. Both models also exhibit larger errors at the interface compared to the bulk, which is a consequence of interfacial environments being underrepresented in the training set for the slab geometry. However, the errors in the ANN-SR model are much larger than the ANN-LR model at the interface, indicating that the multiplicity of long-ranged environments also hinders the training process. Through subsequent active learning steps described below, we are able to optimize both models to have comparable residual force errors below 60 meV/\AA. However,  this procedure does not eliminate the difficulty in using the ANN-SR to describe extended interfaces.
 
\subsection{Active learning steps}
For both reference potentials, we trained a set of 3 distinct ANN-SR and 3 distinct ANN-LR models using a sequence of active learning steps. Data for each of the steps was taken from canonical ensemble simulations of a liquid slab in contact with its vapor. The details of the training of the ANN models depended slightly on the reference potential employed. 

For the dipole model, an initial ANN was constructed using approximately 10000 optimization steps with $N_{\rm batch}=30$. In all cases, the liquid densities and surface tensions predicted by the resulting models were much too low, and several models did not give a stable interface. To remedy these issues, we retrained using an additional dataset generated from multiple molecular dynamics trajectories where $\sigma_{AB}$ was varied between 3.0\AA~and 3.2717\AA, and $k_{\rm bond}$ varied between 20.0\rm kcal/mol \AA~and 100.0 kcal/mol \AA. A total of 724 new configurations were extracted from a combined 400ns of simulation, all having higher densities and longer molecules are typical of the correct reference model. Incorporating these new configurations in the training set with the correct reference-model energies and forces penalises such unphysical geometries and improves the predicted liquid density. In this step the batch size was reduced to 1 for greater training efficiency.

Finally, we retrained the ANN models with an active learning step.  Configurations with a particularly high force error were taken from simulations of the initial models and added to the  training dataset. The loss function was reoptimized, thus improving model predictions for these poorly-trained configurations. Configurations were selected either by the root mean square error across the atomic environments or the maximum error in particular atomic environment. For the ANN-SR models we used a threshold root mean square force error of 7eV/\AA~and maximum error of 14eV/\AA. For the ANN-LR models the thresholds were 0.185eV/\AA~ and 2.0eV/\AA.

%

\begin{figure}
    \includegraphics[width=8.4cm]{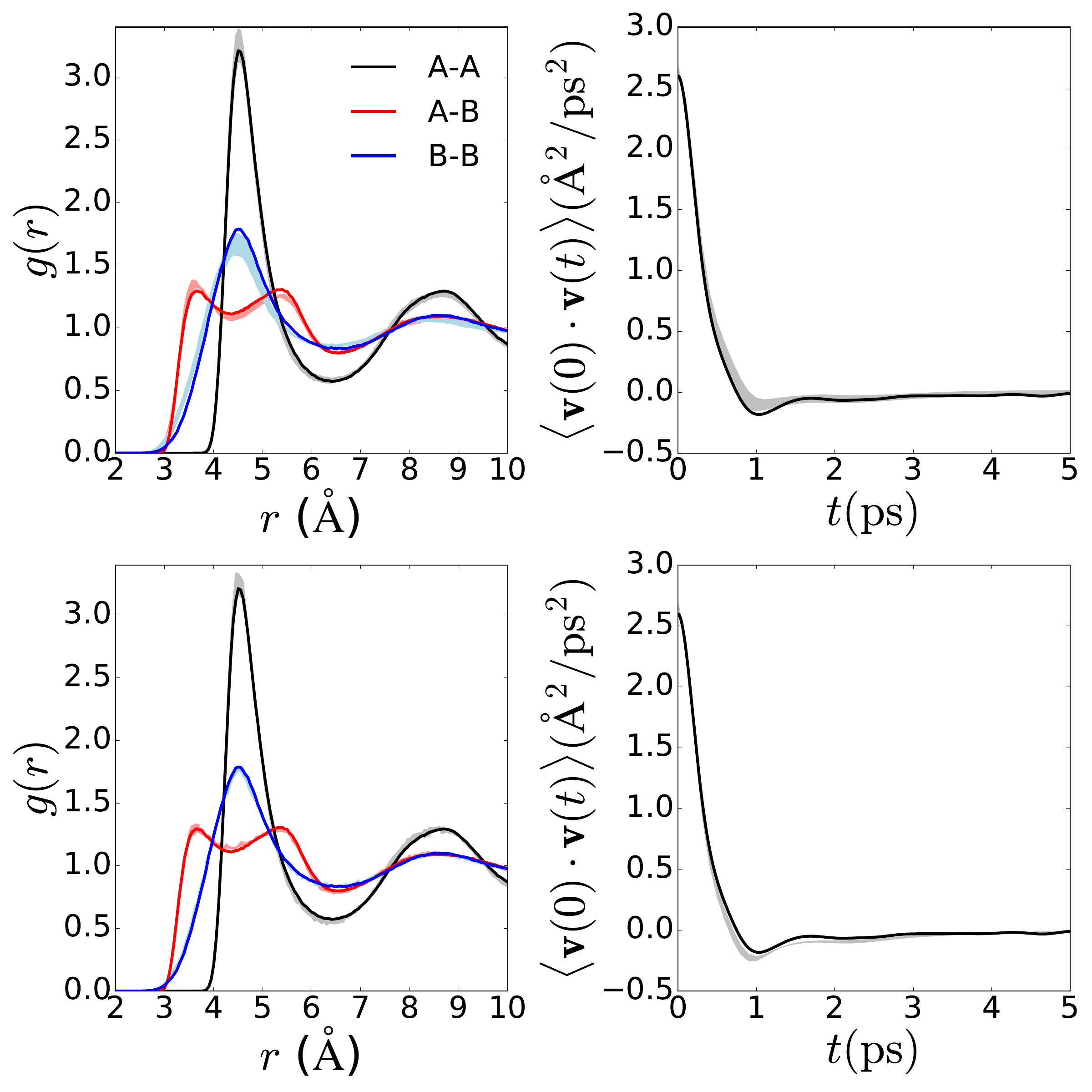}
  \caption{Bulk properties of the dipole fluid. (Top) ANN-SR model predictions of the pair distribution functions (left) and center-of-mass velocity autocorrelation functions. (Bottom) ANN-LR model predictions of the pair distribution functions (left) and velocity autocorrelation functions (right). Solid lines denote the reference force-field calculation and shaded regions denote the range of predictions from 3 ANN models.}
  \label{fig:bulk_properties}
\end{figure}
\begin{figure}
    \includegraphics[width=8.4cm]{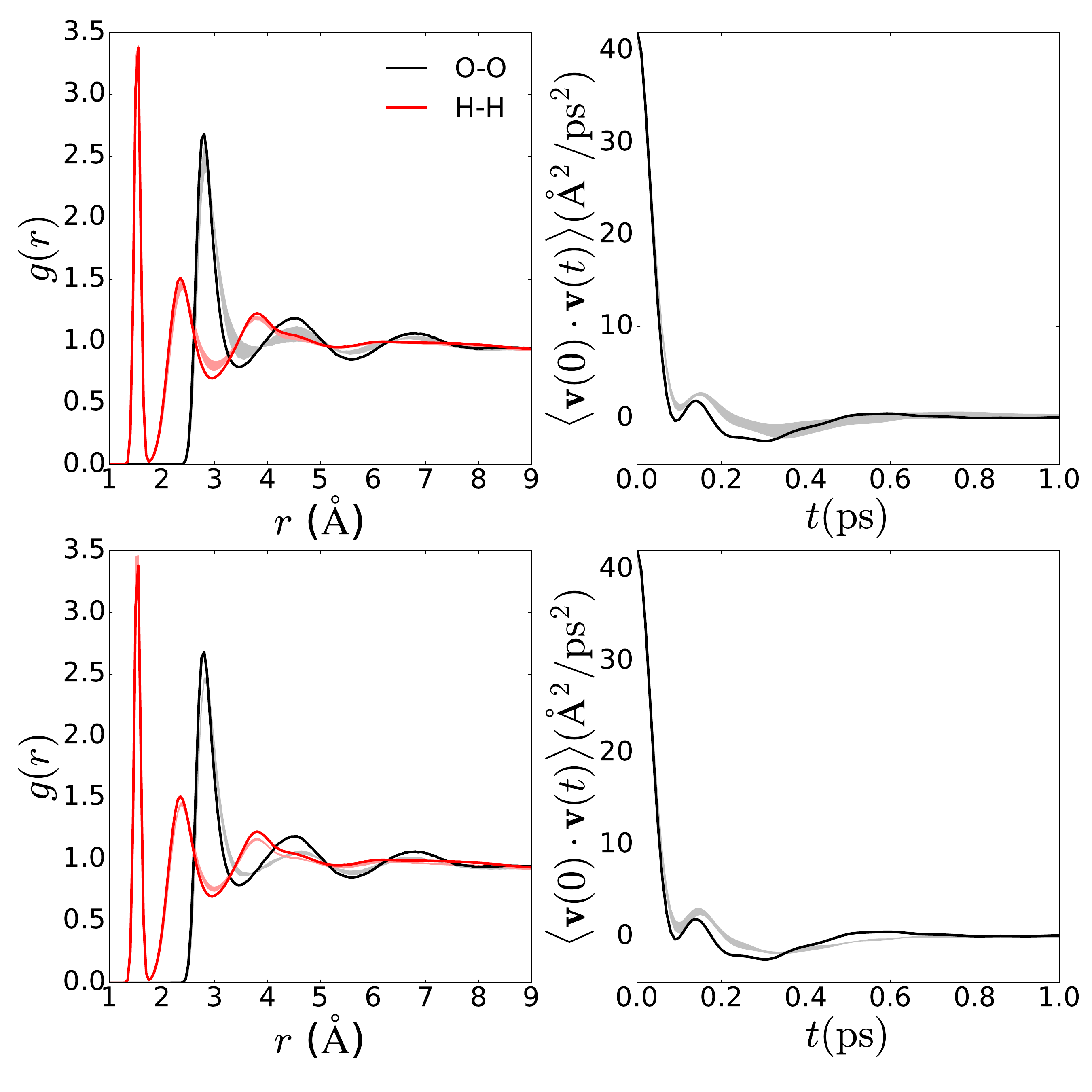}
    \caption{Bulk properties of revPBE-D3 water. (Top) ANN-SR model predictions of the pair distribution functions (left) and center-of-mass velocity autocorrelation functions. (Bottom) ANN-LR model predictions of the pair distribution functions (left) and velocity autocorrelation functions (right). Solid lines denote the reference \emph{ab initio} calculation and shaded regions denote the range of predictions from 3 ANN models.}
  \label{fig:dft_bulk_properties}
\end{figure}

For the water model, an initial ANN was trained using 5000 configurations extracted from a 25 ps AIMD slab simulation and 5000 configurations of bulk water. In this first step, the loss function was optimized to reduce the energy and force error. The initial ANN was then retrained by optimizing the virial error in addition to energy and force errors using 500 configurations, followed by an active learning step where approximately 10000 configurations were added to the previous data set and only energy and force error were optimized. Unlike the dipole model, the reference properties for water configurations are not easy to evaluate, so the active learning configurations were selected using query by committee\cite{krogh1995neural} from five 1 ns trajectories generated by the 3 initial ANN models. Finally, another 2000 configurations from the same initial trajectories were used to optimize the virial error. The final retrained models gave well-correlated force and virial predictions. 

\section{\label{sec:BulkProps}Validation of bulk liquid properties}

Both ANN-SR and ANN-LR achieve consistently small loss functions (below 50meV/\AA) by the end of their active learning cycles, although the latter models generally have smaller errors after an equivalent number of training steps.
 These small loss functions are a necessary but not sufficient condition for reliable predictions of the system thermodynamics or dynamics.
 We first focus on comparing bulk ensemble properties of the ANN models with the reference potential. In a homogeneous liquid, such as the interior of our slab geometry, LMF theory suggests that slowly-varying long-ranged force contributions largely cancel and short-ranged forces dominate the structure of the fluid. We thus expect both ANNs to describe these properties well. Indeed, substantial previous work on ANN models of bulk fluids indicate that such potentials can robustly recover both the thermodynamics and dynamics of their reference systems.\cite{cisneros2016modeling,li2017machine,unke2019physnet,cubuk2017representations,Zhang2018NeurIPS,behler2007generalized,behler2011atom,Wang2018,behler2008pressure,Cheng2018,morawietz2016van,bonati2018silicon,yang2021using,Galib21,kaser2020isomerization,liu2018constructing}

 Figures~\ref{fig:bulk_properties} and \ref{fig:dft_bulk_properties} confirm that both ANN-SR and ANN-LR models are well able to describe the equilibrium structure, as encoded by the pair distribution functions $g(r)$ of the corresponding fluids. Further, both ANN-SR and ANN-LR models are able to describe the equilibrium diffusive dynamics as encoded by the molecular center of mass velocity autocorrelation functions, $\langle \mathbf{v}(0) \cdot \mathbf{v}(t)\rangle$.  Atoms within 10\AA~of the Gibbs dividing surface are excluded from the expectation value of the pair distribution function.
The close agreement between ANN-SR, ANN-LR and benchmark calculations is preserved for both reference systems, and the 
three independently-optimized ANN models for each type, ANN-SR or ANN-LR, are consistent with little variability amongst them. 

\section{Importance of long-ranged interactions for interfacial properties}
Within the context of LMF theory, the inability of the ANN-SR models to describe long-ranged interactions suggests that they should struggle to accurately predict interfacial properties where such interactions become unbalanced. We can anticipate for which properties in particular the ANN-SR model is likely to struggle  by considering a result of LMF theory. The potential decomposition in Eq.~\ref{Eq:LMD} presupposes that the important long-ranged, slowly varying part of the potential is Coulombic in origin. This assumption is reasonable for both the simple dipole fluid and water, where the leading-order asymptotic interactions are dipole-dipole forces. In this case, a truncated model can be constructed that accurately recovers the interfacial properties of the reference system provided an additional external potential. The form of the external potential, denoted $V_R(\mathbf{r})$, is analytically known and given by an electrostatic potential\cite{rodgers2008local}
\begin{equation}
\label{Eq:LMF}
V_R(\mathbf{r}) = \int d\mathbf{r}' \frac{{\rm erf}(\kappa|\mathbf{r}-\mathbf{r}'|)}{4 \pi \varepsilon_0 |\mathbf{r}-\mathbf{r}'|}  \sum_{i} q_i \rho_i(\mathbf{r}')
\end{equation}
where the slowly varying potential ${\rm erf}(\kappa|\mathbf{r}-\mathbf{r}'|)/4 \pi \varepsilon_0 |\mathbf{r}-\mathbf{r}'|$ is convoluted with the charge density $\sum_i q_i \rho_i(\mathbf{r}')$. Note that the atomic densities $\rho_i$ are ensemble averages, in our case obtained directly from the reference potential. For the slab geometry, translational invariance in the plane of the interface renders $V_R(\mathbf{r})$ a function of $z$ only.\cite{cox2020dielectric} 

\begin{figure}
    \includegraphics[width=8.4cm]{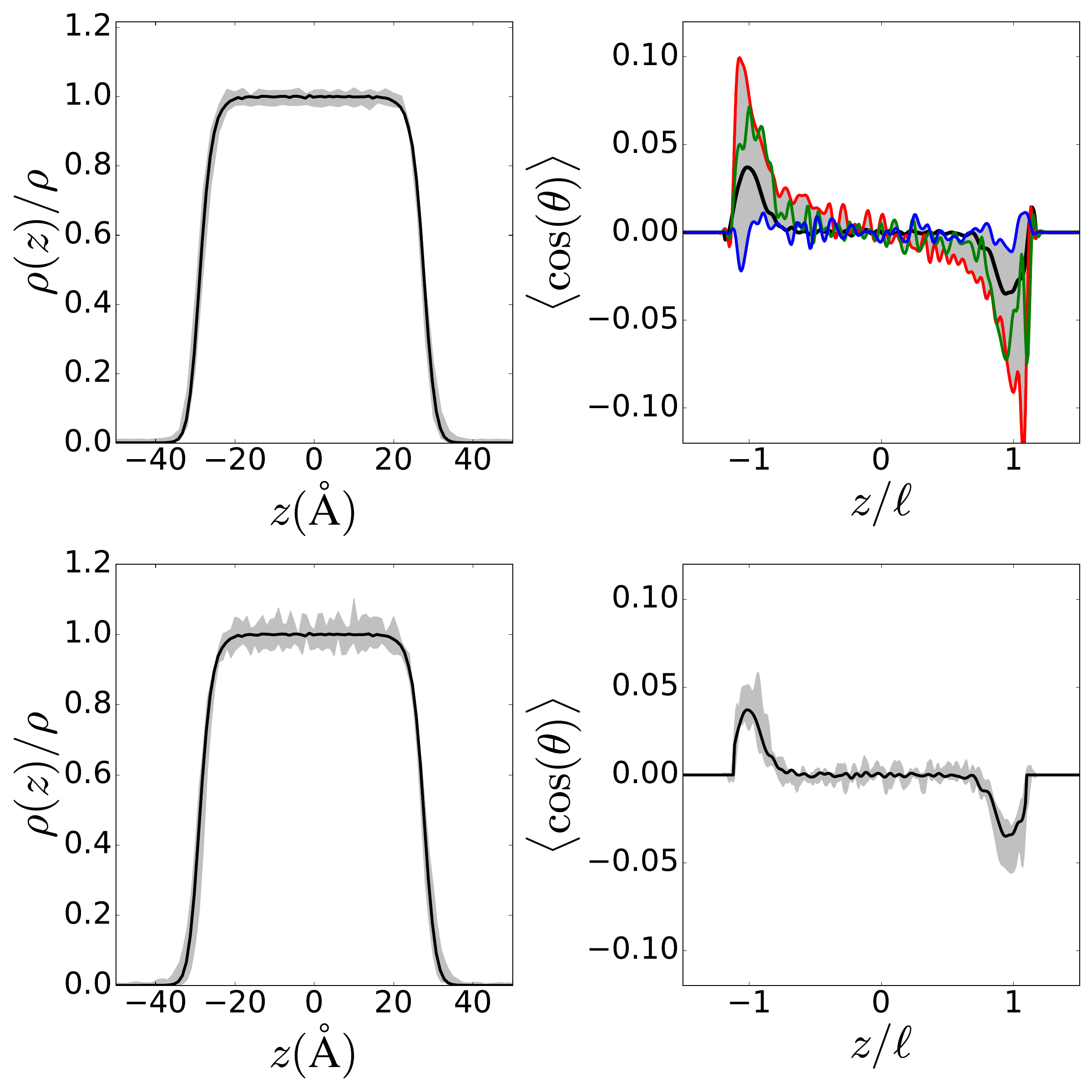}
  \caption{Interface properties of the dipolar fluid. (Top) ANN-SR model predictions of the A-site density (left) and orientational order (right). (Bottom) ANN-LR model predictions of the A-site  density (left) and orientational order (right). Solid black lines denote the reference forcefield calculation and shaded regions denote the range of predictions from 3 ANNs. The different colored lines indicate different ANN-SR models. In the orientational order plots, $z$ bins with a density less than 2\% of the bulk liquid value are not shown.}
  \label{fig:int_properties}
\end{figure}

\begin{figure}
    \includegraphics[width=8.4cm]{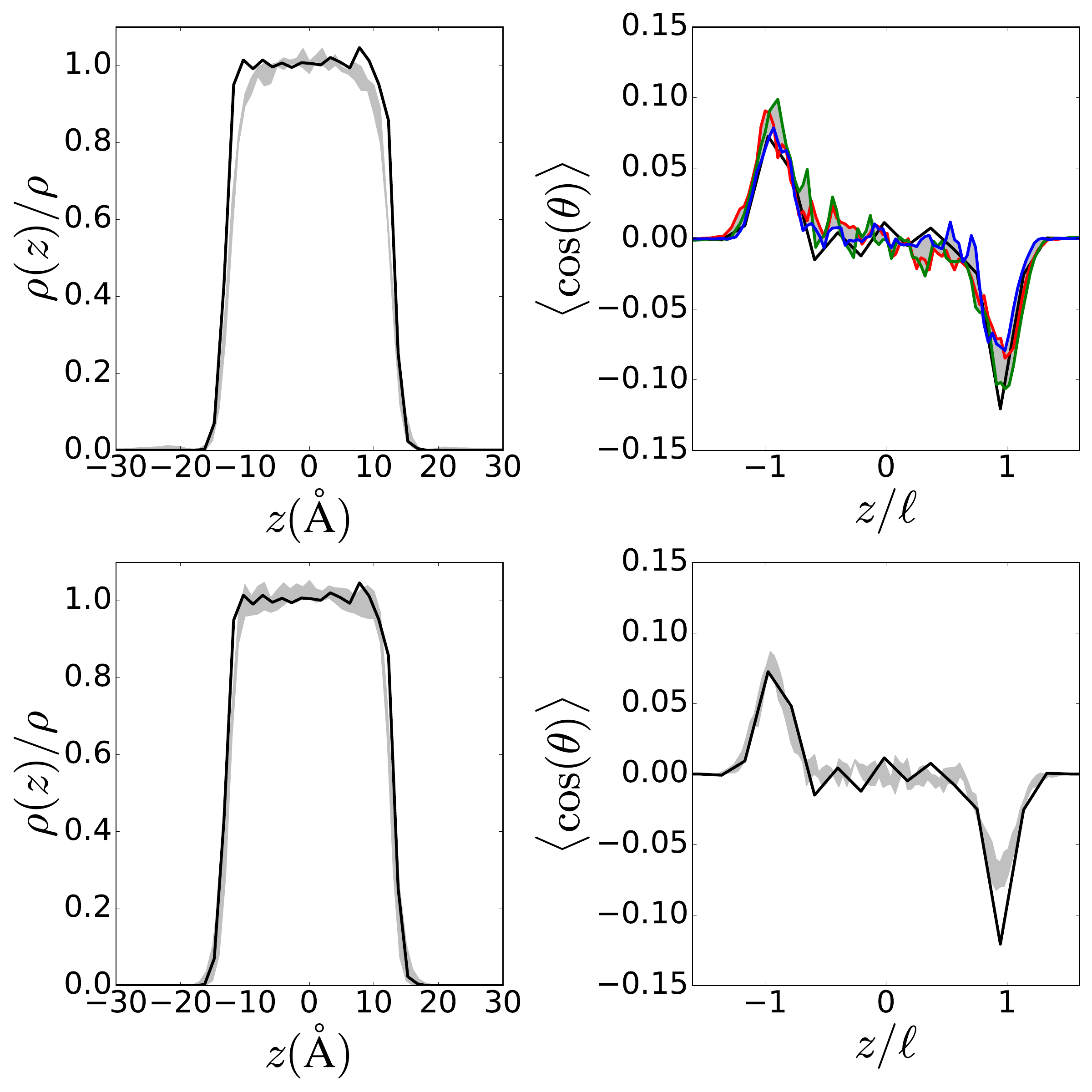}
  \caption{
  Interface properties of revPBE-D3 water. (Top) ANN-SR model predictions of the density (left) and orientational order (right). (Bottom) ANN-LR model predictions of the density (left) and orientational order (right). Solid black lines denote the reference \emph{ab initio} calculation and shaded regions denote the range of predictions from 3 ANNs. The different colored lines indicate different ANN-SR models.
  }
  \label{fig:int_properties_dft}
\end{figure}

Since $V_R(\mathbf{r})$ is an external electrostatic potential, properties related to spatial or thermal fluctuations of the collective polarization will be most affected by the absence of long-ranged interactions. Properties weakly correlated with an electrostatic potential are likely to be well described by a model with purely short-ranged interactions, like the ANN-SR models. Figures~\ref{fig:int_properties} and \ref{fig:int_properties_dft} illustrate the number density and orientational order profiles relative to the liquid-vapor interfaces of our slab geometries. Consistent with the expectation that density fluctuations are weakly correlated with polarization fluctuations,\cite{remsing2013dissecting} we find that both the ANN-SR and ANN-LR are able to reasonably reproduce the sigmoidal density profiles, $\rho(z)/\rho$, of the reference dipolar fluid and \emph{ab initio} water models they have been trained on. This agreement can be understood as the weak response of the mass density of neutral molecules to an external electric field. The bulk liquid density $\rho$ for the dipolar fluid model is $\rho=0.036$\AA$^{-3}$ while that for the revPBE-D3 water model $\rho=0.96\mathrm{g/cm}^3$. The value of the water density is in reasonable agreement with recent estimates,\cite{galib2017mass} which found a strong dependence on the basis set. Further, as we observed for the bulk liquid properties, there is only a small variance between the set of 3 independently trained potentials in each class of model.

The orientational order profiles shown in Figures~\ref{fig:int_properties} and \ref{fig:int_properties_dft}, are computed by evaluating the angle, $\theta$, between the molecule's dipole and the $z$ axis. For both the dipolar fluid and the \emph{ab initio} water model, the orientational order profiles are generally poorly predicted by the ANN-SR models, but well recovered by the ANN-LR models. This difference is rationalized within the framework of LMF theory as the molecular orientation is responsible for spatial fluctuations of polarization in the fluid, which are therefore sensitive to the unbalanced electrostatic potential generated by truncating the dipolar interactions.\cite{cox2020dielectric}  For both the dipolar fluid and water, we find a variety of behaviors in the ANN-SR models, including models with relatively little orientational alignment throughout the slab, and models with persistent alignment permeating throughout even the bulk of the liquid. This persistent alignment, seen in ANN-SR models trained on both the dipolar fluid and the water but not for any ANN-LR model, has also been observed recently in truncated forcefield models of water.\cite{cox2020dielectric} As with the bulk properties, the ANN-LR models exhibit little variability within the set of 3 independently trained potentials.

Aside from structural measures of the interface, we have also considered the impact of long-ranged interactions on the liquid-vapor surface tension, $\gamma$. We computed the surface tension using the approach by Kirkwood, employing a difference in the diagonal components of the virial,\cite{frenkel2001understanding}
\begin{equation}
\gamma = \left \langle \Xi_{zz}-\frac{1}{2}\left (\Xi_{xx}+\Xi_{yy} \right ) \right  \rangle \frac{L_z}{2}
\end{equation}
where $\Xi_{ii}$ is the $i$th component of the virial matrix defined previously.  Previous reports have indicated that short-ranged ANN models can accurately predict the surface tension.\cite{Wohlfahrt2020} We find that this is indeed the case, with the reference surface tension value computed by the empirical forcefield dipolar fluid model falling within the range of both ANN-SR and ANN-LR models. The value of $\gamma$ for revPBE-D3 water is not known, since it is too expensive to evaluate directly, but previous ANN results suggest that it is 68 mN/m at 300 K.\cite{Wohlfahrt2020} Our ANN-SR estimates are slightly lower than this value. Unlike the orientation, which will depend linearly on an external field generated by the truncated interactions, the leading order dependence of the surface tension is quadratic. This is because the slab has inversion symmetry and cannot depend on the sign of the applied field. Thus the agreement between the reference calculations and the ANN-SR models suggests that the majority of $\gamma$ is determined locally, while the remainder is small enough that its quadratic correction is negligible.

\begin{table}
  \begin{tabularx}{.45\textwidth}{c|c|c}
      \quad Model \quad& \quad $\gamma_{\rm Dipole}$ (mN/m) &\quad $\gamma_{\rm H_2O}$ (mN/m) \\
  \hline 
    Reference\quad & \quad 4.5 $\pm$ 0.3 &  68\cite{Wohlfahrt2020} \\
   ANN-SR & \quad 5.2 $\pm$ 0.8 & \quad 54 $\pm$ 6  \\
    ANN-LR & \quad 5.0 $\pm$ 0.4 & \quad  66 $\pm$ 14 
  \end{tabularx}
  \caption{\label{tab:gamma}Surface tension values for different neural network models parameterized for the dipole forcefield, $\gamma_{\rm Dipole}$ and the \emph{ab initio} water, $\gamma_{\rm H_2O}$. The error in the reference is statistical, and those of the ANN models are calculated from the range of 3 models, which is larger than the statistical uncertainty. 
  }
\end{table}

\section{ANN-SR approximation of the LMF potential}
The reference training data includes force contributions from atoms outside the symmetry function cutoff for the ANN-SR models. The ANN still attempts to fit these contributions to its symmetry functions, so a natural question is whether the ANN attempts to learn the LMF potential or whether it overfits unrepresentable forces to particular atomic environments. Such overfitting would likely lead to a high extrapolation error for the model. The orientational profiles in Figs.~\ref{fig:int_properties} and \ref{fig:int_properties_dft} suggest that some ANN-SR models learn some degree of the long-ranged forces, since not all of them generate a net polarization in the bulk-like region. Also the ANN-SR models produce a smaller polarization at the interface than do truncated models without an added explicit long-ranged potential.\cite{cox2020dielectric} This argument is elaborated further in Appendix B.

Figure~\ref{fig:LR_forces} compares an estimate of the implicit long-ranged component in ANN-SR potentials with an analytical approximation of the long-ranged force in the reference dipole model. Specifically, we evaluate the LMF potential by solving Eq.~\ref{Eq:LMF} using the charge density computed from reference forcefield calculations, and take its numerical gradient in the $z$ direction. This is the one-shot approach to solving the self-consistent LMF equations.\cite{rodgers2008local,cox2020dielectric} To estimate the long-ranged forces in an ANN-SR model, we assume that the short-ranged components are well approximated by the short-ranged part of the reference potential, $U_{\rm SR}$, defined in Eq.\ref{Eq:LMD}. The residual quantity $F_z(\mathbf{R})+\partial_z U_{\rm SR}(\mathbf{R})$ therefore estimates the implicit contribution of distant atoms to the force predicted by an ANN-SR model. The ensemble average of this residual force depends slightly on the pair distribution of the interacting atoms, whereas the LMF force is an external field independent of correlations between atoms. However, at long distances the pair distribution is almost uniform, and we have confirmed that the discrepancy between the LMF force and the reference $\langle F_z+\partial_z U_{\rm SR}\rangle$ is small.

Figure~\ref{fig:LR_forces} shows that long-ranged forces in ANN-SR models resemble the LMF potential but do not exactly recover it. This resemblance indicates that the ANN-SR models attempt to approximate a molecular field potential, which is surprising given that the scale separation in the reference data is implicit. However, the attempt is not particularly successful and there is significant variation in the scale and shape of the implicit long-ranged forces as a result. While some of the models reproduce the results of LMF theory, most over- or under-estimate the restructuring potential significantly. It is possible that tailoring loss functions to recover this LMF potential may be successful for specific systems and geometries, however we expect that the implicitly-learned long-ranged force will be generally poorly transferable given that the magnitude of the LMF potential depends on the system geometry.

For comparison, in Fig.~\ref{fig:LR_forces} we also show the residual force computed for an ANN-LR model. By construction, the long-ranged component of $F_z$ in this model exactly matches the reference and so the agreement with the LMF force is much better than for the ANN-SR models. This agreement underscores the fundamental improvement in interfacial representation with the ANN-LR approach. The small discrepancy between the ANN-LR and LMF forces quantify the deviation that may be expected from the presence of persistent pair correlations discussed above, and errors in the ANN short-ranged force relative to $U_{\rm SR}$. The difference between the ANN-SR models and the LMF curve is much too large to be explained by any of these factors, illustrating the failure of the truncated symmetry functions to capture interfacial interactions.

\begin{figure}[t]
  \includegraphics[width=8.5cm]{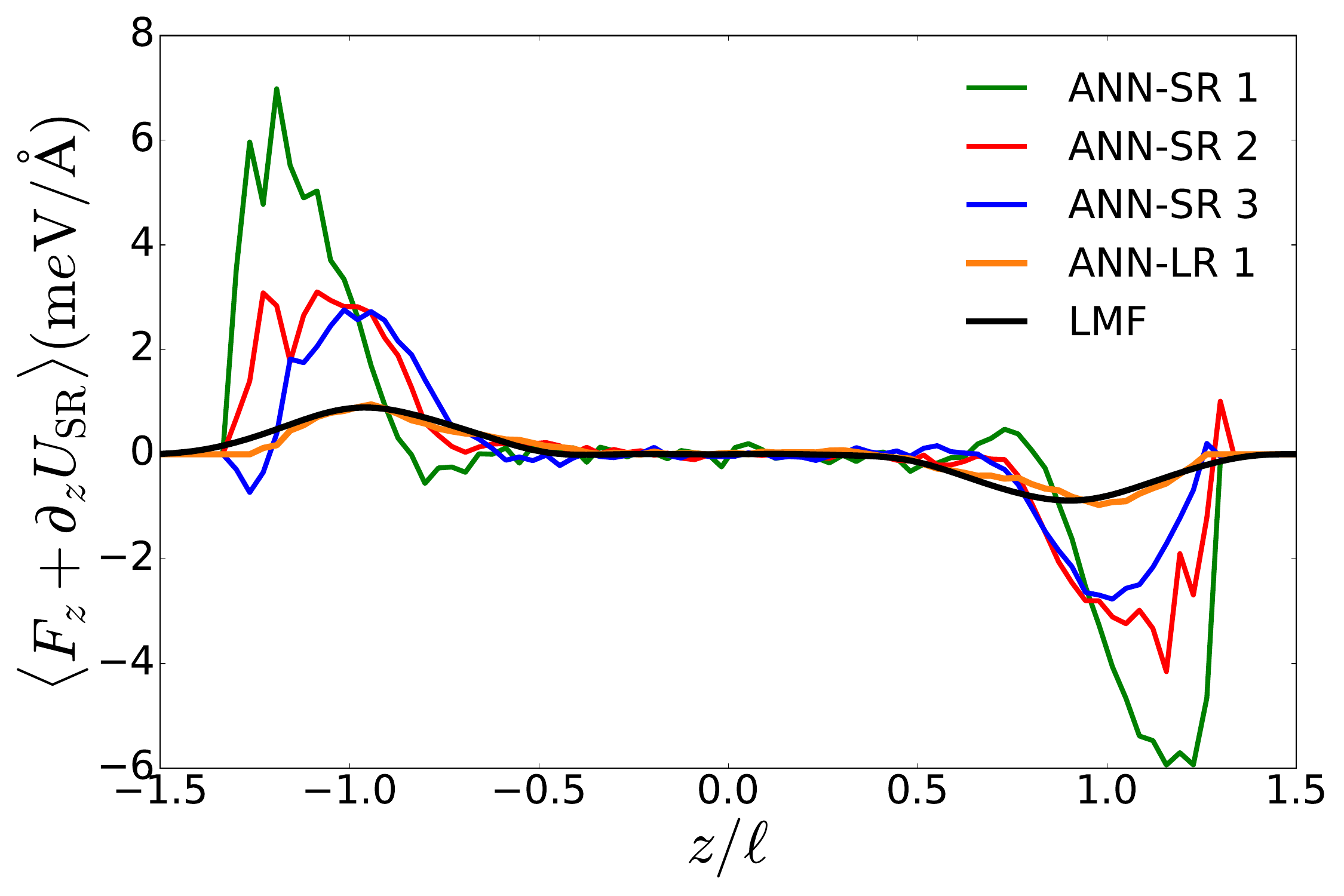}
  \caption{Estimates of the long-ranged force learned implicitly by ANN-SR potentials and one typical ANN-LR model, compared to the LMF force on an A-type site. For the four ANN models, the ensemble average is performed over a long trajectory generated by the respective model, and averaged over all A-sites within each $z$ interval. Only $z$ values where $\rho(z)/\rho>0.02$ are shown.}
  \label{fig:LR_forces}
\end{figure}

\section{Sensitivity of ANN-LR to the charge model}
The preceding section shows that incorporating explicit long-ranged forces, for example through an ANN-LR approach, is necessary to reliably capture the correct interfacial behavior using artificial neural network forcefields. 
Constructing an ANN-LR model of \emph{ab initio} water requires a model for the unknown charge distribution in this system.
The choice of this model should not affect the short-ranged properties (at least in the limit of perfect training) because the long-ranged force subtracted in Eq.~\ref{Eq:LMD} is added back to the ANN-LR model during simulations.
However, collective fluctuations of molecular orientation are sensitive to the exact long-ranged force component. It is therefore important to explore the sensitivity, or insensitivity, of our results to our specific choice of long-ranged model, which we have done by varying the charges selected in Eq.~\ref{Eq:LMD}.

Specifically, we trained three sets of 5 ANN-LR models using the previously-described active leaning procedure, each set having a different magnitude of charge on the hydrogen, $q^*_H/q_e = 0.3814, 0.4238, 0.4662 $, where $q^*_H = 0.4238q_e$ is the SPC/E model value. The modified charges correspond to differences of $\pm 10\%$. Oxygen charges were also adjusted to preserve neutrality of each molecule.
We first confirmed that the pair distribution functions and velocity auto-correlation functions resulting from the charge-modified models all fall within the spread of predictions from the original ANN-LR models (trained on SPC/E charges). This agreement indicates little sensitivity of the bulk properties to the long-ranged force model, as expected from liquid-state theory and the results of previous sections.

Its less clear \emph{a priori} whether the interfacial properties are insensitive to the long ranged model employed. In order to quantify this dependence, we study the effect of modifying the charge model on the density and orientational profiles. Specifically, we fit the density profile to a functional form 
\begin{equation}
\rho(z) = \rho \left ( \frac{1}{1+e^{-(z+\ell)/w_\rho}}- \frac{1}{1+e^{-(z-\ell)/w_\rho}} \right )
\end{equation}
with shape descriptors $\rho$ for the bulk density and $w_\rho$ for the width of the liquid-vapor interface. Additionally we fit the orientational profile to the functional form
\begin{equation}
\langle \cos(\theta) \rangle = \alpha \left ( e^{-(z+\ell)^2/ w_\theta^2}- e^{-(z-\ell)^2/ w_\theta^2}  \right )
\end{equation}
with shape descriptors $\alpha$ for the amplitude of the polarization and $w_\theta$ for the width of the polarized region at the interface. Both functional forms provide good fits across the range of ANN-LR models constructed. 

\begin{figure}[t]
  \includegraphics[width=8.5cm]{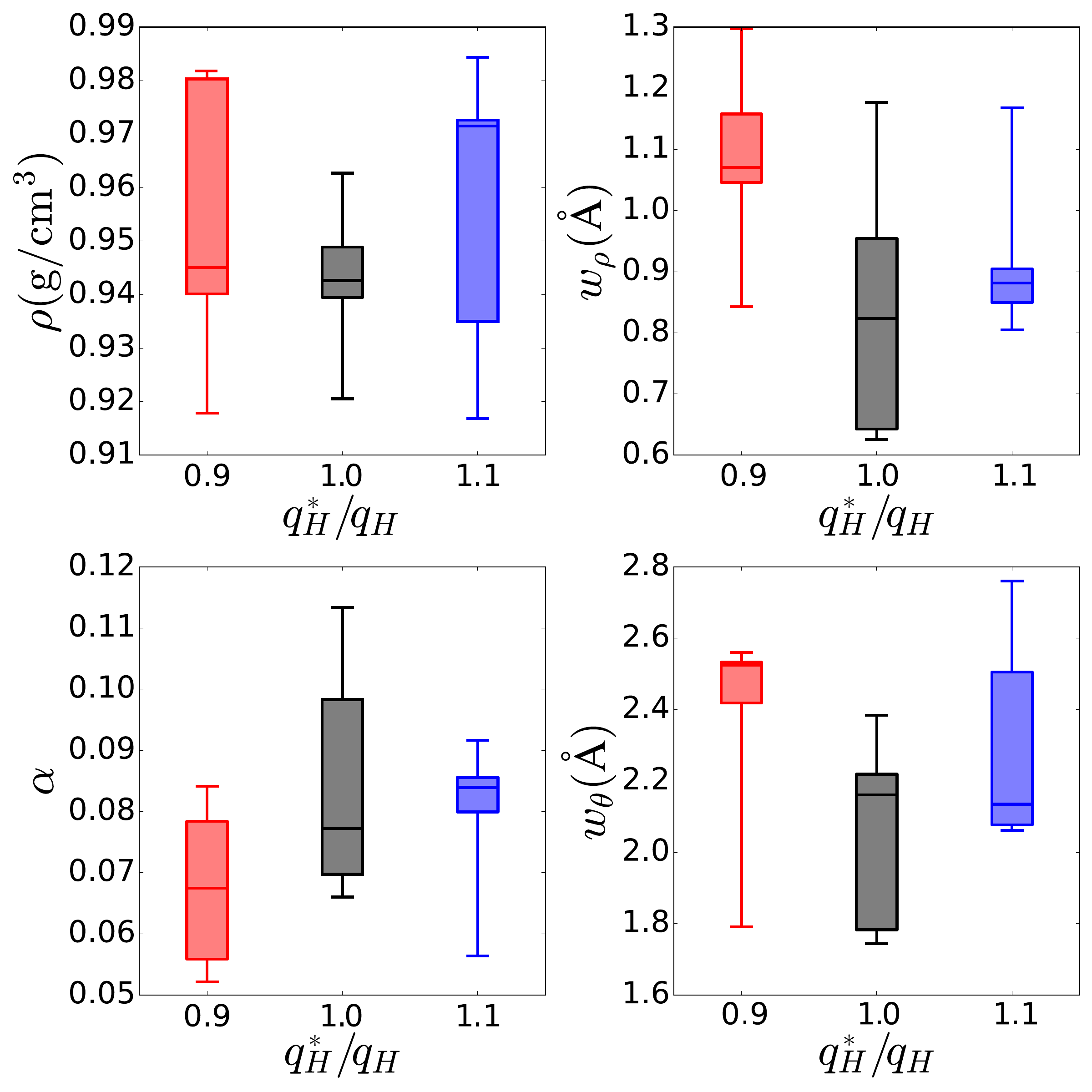}
  \caption{Box plots of the shape parameters for the density (top) and orientation (bottom) profiles denoting the median, first quartile and range over 5 independently trained ANN-LR with varying fixed charge models for a slab of revPBE-D3 water.}
  \label{Fig:Stat}
\end{figure}

Summary statistics for $\rho,\alpha,w_\rho$ and $w_\theta$ for the 15 charge-varied models are shown in Fig.~\ref{Fig:Stat}. For all of the shape descriptors, the variability  within a set of models with the same charge is larger than the variation between models with different charges. The two width parameters vary on sub-angstrom scales, with the typical values of $w_\rho \approx 0.9$\AA \, and $w_\theta \approx 2.2$\AA. There is no systematic trend in the amplitude of the orientational order $\alpha$, though we do find that the bulk density is more variable between models for the lower and higher charge cases than for the SPC/E charges employed.  We are left with the conclusion that the internal variability of equivalent ANN models with different stochastic training histories is a much larger effect than the variation between different long-ranged force models. Thus, the ANN-LR models are relatively insensitive to the specific charges employed.

\section{Conclusions}
ANN models are capable of describing complex, many-bodied interactions, enabling their deployment in a wide range of chemical phenomena. Currently, ANN potentials are overwhelmingly formulated with atomic descriptors that are local and short-ranged. Here, we have used LMF theory as a framework to understand in what cases this locality can offer a good approximation to equilibrium structure and dynamics, and when we should expect the neglect of long-ranged interactions to matter. In systems with translational invariance, long-ranged slowly varying forces like those between two molecules with permanent  dipoles in a bulk fluid largely cancel. As such the neglect of long-ranged forces is inconsequential. However, at extended interfaces like that between a liquid and its vapor, truncated interactions result in unbalanced forces that can polarize and distort local molecular structure. While in principle, ANNs may be able to uncover an approximation to an effective external field that can correct for these unbalanced interactions, we have found that in practice this approximation is difficult to achieve consistently. Even well-trained models that describe properties of the bulk liquid accurately will usually predict overly-polarized interfaces and unphysical net polarization of molecules in the bulk-like interior of the liquid.

To mediate the failings of truncated ANNs, we have proposed training models only on the short-ranged interactions and employing a model of the long-ranged forces. Rather than training models with a full electrostatic force model subtracted from them, we use insights from LMF theory to subtract only the part traditionally evaluated in reciprocal space. This approach leads to models with a natural separation of length scales, which are able to reproduce reference models based on an empirical fixed charge potential as well as one derived from density functional theory.  This separation has the added benefit that local interactions are done completely by the neural network, while long-ranged interactions are done efficiently through traditional Ewald summations, avoiding the need to sum independent local interactions.  We have employed simple models for the long ranged interactions based on fixed point charges, with the expectation that higher multipoles decay faster with distance and are accommodated by the short-ranged ANN. Further, the local description of the ANN employed here is flexible enough to generate models that are largely insensitive to the specific charge model employed. With an ability to study extended interfaces with ANNs robustly a wide range of phenomena become tractable, including understanding the effects of extended interfaces on chemical reactivity.\cite{niblett2021ion,schile2019rate,kattirtzi2017microscopic,benjamin2015reaction,mundy2006first,lewis2011does} Addressing such questions will undoubtedly advance fields such as atmospheric chemistry and catalysis where strongly inhomogeneous systems are ubiquitous.

\section*{DATA AVAILABILITY}
The data that support the findings of this study are available from the corresponding author upon reasonable request.

\section*{Acknowledgements}
We thank Stephen Cox for useful discussions on LMF theory. This material is based on work supported by the U.S. Department of Energy, Office of Science, Office of Advanced Scientific Computing Research, Scientific Discovery through Advanced
Computing (SciDAC) program, under Award No. DE-AC02-05CH11231. This research used resources of the National Energy Research Scientific Computing Center (NERSC), a U.S. Department
of Energy Office of Science User Facility operated under Contract No. DE-AC02-05CH11231.

\section*{\label{sec:AppImplementation}Appendix A: ANN-LR implementation in LAMMPS}

We implemented the ANN-LR procedure by using the DeePMD LAMMPS interface\cite{Wang2018} to compute short-ranged ANN energies in combination with the PPPM reciprocal-space solver from the KSPACE package.
For the water model, we obtained short-ranged energy and force training data by subtracting model reciprocal-space forces, given by the PPPM solver, from the {\it ab initio} DFT values. For the dipole model, we obtained training data by computing the short-ranged part of the forcefield directly with the Gaussian Truncated pair potential.\cite{cox2020dielectric}

Recall that in the reference model for the dipolar fluid, bonded atoms interact only through a harmonic bonding potential. However, the PPPM solver does not discriminate between bonded and non-bonded atom pairs and so adds a reciprocal-space coulombic force between the two sites of each dimer. In normal LAMMPS usage the short-ranged force calculation would remove this force, but the DeePMD interface does not, so our ANN-LR model contains a small intramolecular contribution not present in the reference model.

However, we have found that the magnitude of this force is never greater than 20\% of the harmonic bonding potential, and is typically much smaller. Moreover, the reciprocal-space force depends only weakly on the atomic separation whereas the harmonic restoring force varies rapidly, so the error becomes increasingly negligible away from the equilibrium bond length. We have found no significant effect from this error on either the distribution of bond lengths or the intermolecular structure (see sec.~\ref{sec:BulkProps}).

We note further that in a more realistic ANN problem such as the {\it ab initio} water model, intramolecular coulomb forces would likely be included in the reference potential and this problem would not arise.

\begin{figure}[t!]
  \includegraphics[width=8.5cm]{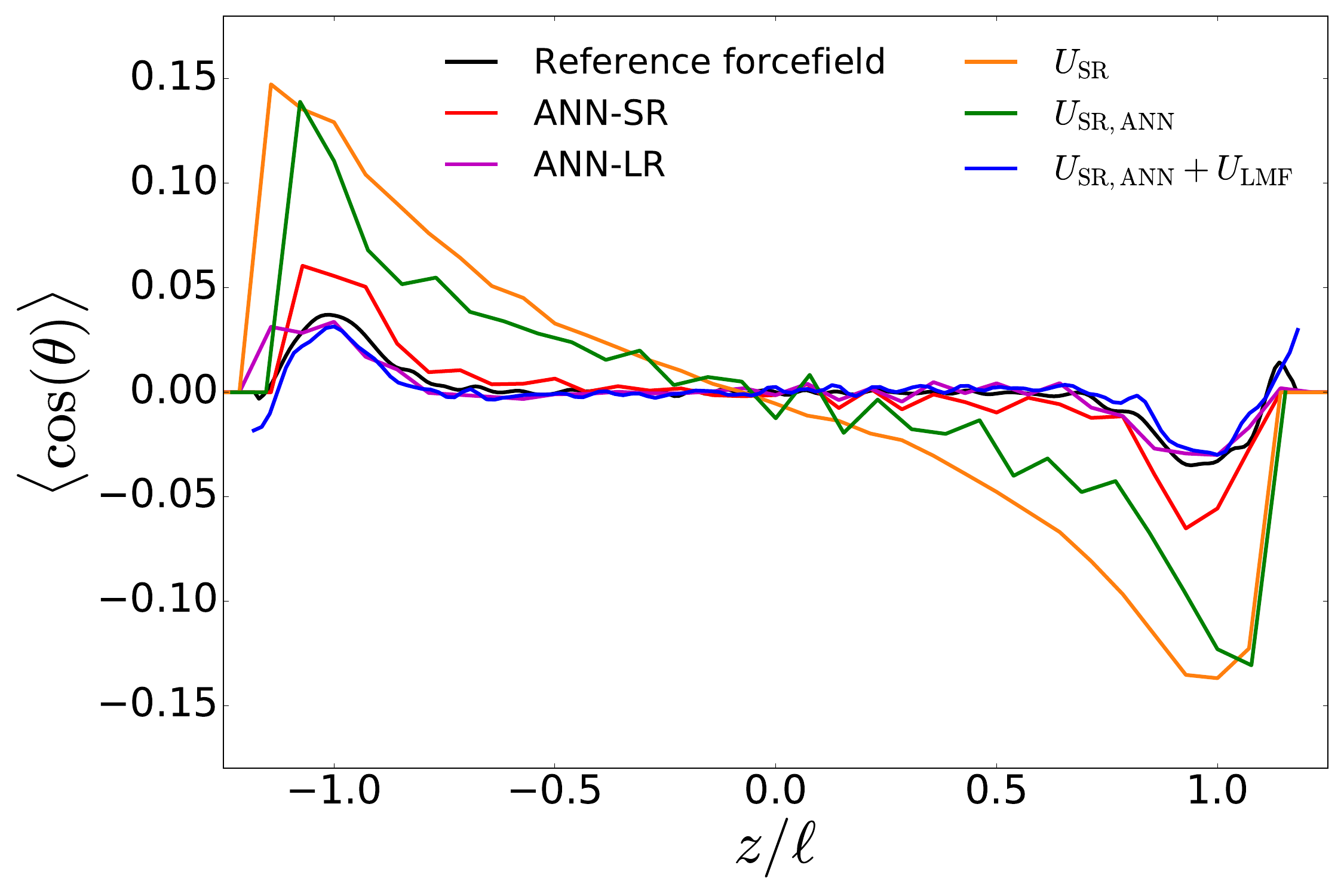}
  \caption{Orientational profiles for several models of a dipolar fluid. In all cases, the $z$ coordinates are scaled to place the mean Gibbs interfaces at $z=\pm\ell$. $z$ bins with a density less than 2\% of the bulk liquid value are not shown.}
  \label{Fig:AllPolarisation}
\end{figure}
\section*{\label{sec:AppOrientations}Appendix B: The Importance of Long-Ranged Forces to Reproduce Orientational Order}

Figure~\ref{Fig:AllPolarisation} demonstrates how different representations of long-ranged forces can result in very different molecular polarizations near a liquid-vapor interface.
The reference forcefield, ANN-SR and ANN-LR labels all have the same meaning as in the main text. Only the best-performing ANN-SR and ANN-LR models are shown. $U_{\rm SR}$ indicates the short-ranged reference forcefield defined in eq.~\ref{Eq:LMD} and $U_{\rm SR,ANN}$ is a neural network trained to reproduce only that short-ranged force. Finally, $U_{\rm SR,ANN}+U_{\rm LMF}$ indicates a force-field that combines the reference short-ranged force with the restructuring potential defined in eq.~\ref{Eq:LMF}. 

The models depicted fall into three groups with qualitatively different orientational orders. Models with no long-ranged components, $U_{\rm SR}$ and $U_{\rm SR,ANN}$, exhibit large interfacial polarization and a continuous polarization gradient throughout the liquid slab. The reference and ANN-LR models which have exact long-ranged forces, and the LMF model which approximates these forces by a constant external field, all lack orientational order within the slab and all predict the same magnitude of polarization at the interface. The contrast between these two groups illustrates the importance of long-ranged forces to capture polarization fluctuations.

The final group consists of the ANN-SR model, which falls in between the other two groups: the interface is over-polarized relative to the reference but much less so than the  $U_{\rm SR}$ model, and a small polarization gradient persists in the interior of the liquid slab.
This intermediate behavior demonstrates that the short-ranged symmetry functions used in the neural network representation are capable of implicitly learning some degree of long-ranged interactions, even though they cannot represent these forces exactly.

\section*{References}
\bibliography{interfaces}
\end{document}